\documentclass[12pt,draftclsnofoot,journal,onecolumn]{IEEEtran}

\usepackage[english]{babel}
\usepackage{amsmath,amssymb}
\usepackage{times}
\usepackage{relsize}
\usepackage{graphicx}
\usepackage{url}
\usepackage{booktabs}
\usepackage{multirow}
\usepackage{mparhack}
\usepackage{subfigure}
\usepackage{authblk}
\usepackage{amsthm}

\usepackage[noend]{algorithmic}
\usepackage[linesnumbered,ruled,vlined]{algorithm2e}

\usepackage{array}
\usepackage{cite}
\usepackage{eqparbox}
\usepackage{mdwmath}
\usepackage{balance}
\usepackage{epsfig}
\usepackage{xcolor}

\usepackage{mathtools}
\usepackage{bm}
\usepackage{amsfonts}
\usepackage[all]{xy}
\usepackage{etoolbox}
\usepackage{graphicx}
\DeclareMathAlphabet{\mathantt}{OT1}{antt}{li}{it}
\DeclareMathAlphabet{\mathpzc}{OT1}{pzc}{m}{it}

\usepackage{accents}

\DeclareFontFamily{OT1}{pzc}{}
\DeclareFontShape{OT1}{pzc}{m}{it}%
  {<-> s * [1.1] pzcmi7t}{}
\DeclareMathAlphabet{\mathpzc}{OT1}{pzc}%
                     {m}{it}

\DeclareMathOperator{\argmin}{\arg\min}

\makeatletter
\patchcmd{\@maketitle}
  {\addvspace{0.5\baselineskip}\egroup}
  {\addvspace{-1.45\baselineskip}\egroup}
  {}
  {}
\makeatother

\title{Cost-Optimal Caching for D2D Networks with User Mobility: Modeling, Analysis, and Computational Approaches}

\author{
Tao~Deng,~\IEEEmembership{Student Member,~IEEE},
Ghafour~Ahani,
Pingzhi~Fan,~\IEEEmembership{Fellow,~IEEE}, and Di~Yuan,~\IEEEmembership{Senior Member,~IEEE}
\thanks{The paper is a significant extension of a previous work submitted to IEEE Globecom \cite{TDeng2017Cost}.} 
\thanks{T.\ Deng and P.\ Fan  are with the School of Information Science and Technology, Southwest Jiaotong University,
Chengdu, Sichuan 610031, China (e-mail: dengtaoswjtu@foxmail.com and p.fan@ieee.org). }%
\thanks{G.\ Ahani and D.\ Yuan are with the Department of Information Technology, Uppsala University, 751 05 Uppsala, Sweden (e-mail:
\{ghafour.ahani, di.yuan\}@it.uu.se)}

}

\begin{document}

\maketitle

\begin{abstract}
Caching popular files at user equipments (UEs) provides an effective way to alleviate the burden of the backhaul networks.
Generally, popularity-based caching is not a system-wide optimal strategy, especially for user mobility scenarios.
Motivated by this observation, we consider optimal caching with presence of mobility. 
A cost-optimal caching problem (COCP) for device-to-device (D2D) networks is modelled, in which the impact of user
mobility, cache size, and total number of encoded segments are all accounted for.
Compared with the related studies, our investigation guarantees that the collected segments are non-overlapping, takes into account the
cost of downloading from the network, and provides a rigorous problem complexity analysis.
The hardness of the problem is proved via a reduction from the satisfiability problem. Next, a lower-bounding function of the objective
 function is derived. By the function, an approximation of COCP (ACOCP) achieving linearization is obtained, which features two advantages.
First, the ACOCP approach can use an off-the-shelf integer linear programming algorithm to obtain the global optimal solution, and it can effectively deliver solutions for small-scale and medium-scale system scenarios.
Second, and more importantly, based on the ACOCP approach, one can derive the lower bound of global optimum of COCP, thus enabling performance benchmarking of any sub-optimal algorithm.
To tackle large scenarios with low complexity, we first prove that the optimal caching placement of one user, giving other users'
caching placements, can be derived in polynomial time.
Then, based on this proof, a mobility aware user-by-user (MAUU) algorithm is developed.
Simulation results verify the effectivenesses of the two approaches by comparing them to the lower bound of global optimum and conventional caching
 algorithms.
\end{abstract}

\IEEEpeerreviewmaketitle
\section{Introduction}
\subsection{Motivations}
With rapid emergence of new services and application scenarios, such as social networks (e.g., Twitter and Facebook), multimedia contents (e.g., YouTube), and Internet of things (IoT) etc., explosive growth in mobile data traffic and massive device connectivity are becoming two main challenges for existing cellular networks. Hyper-dense small cell networks have been recognized as a promising technology to achieve higher network capacity in fifth-generation (5G) wireless networks \cite{JGAdrews2014,IHwang2013}. However, due to a large number of connections between the base stations (BSs) and core network (CN), the backhaul networks will face a heavy burden \cite{XGe20145G}, calling for research from both the academia and industry. Caching is a promising technology to alleviate the burden of the backhaul networks by storing the required files or contents in advance at the edge devices \cite{Fan2016Coping,Golrezaei2012Femtocell,Peng2016Fog}, e.g., small cells and user equipments (UEs).

With caching, users can obtain their requested files from the edge devices so as to improve the network performance in terms of energy efficiency and file downloaded delay, and at the same time reduce the burden of backhaul \cite{XWang2014Cache}. The caching performance depends heavily on the cache placement strategy.
Although the conventional strategy of caching popular files can improve the probability that the users will find the files of interest in their local caches, it is not a system-wide optimal solution, especially for user mobility scenarios.
Therefore, it is necessary to revisit the caching problem with user mobility and investigate the following questions:
  \begin{itemize}
\item
How to make the best use of user mobility to design approaches for optimizing content caching?
\item
How much will mobility help?
\end{itemize}
To address the two questions,
we consider caching at mobile users and investigate cost-optimal caching for device-to-device (D2D) networks. More specifically, the inter-contact model is used to describe the mobility pattern of mobile users.
The mobile users can collect segments of files when they meet each other. If the total number of collected data segments is not enough to recover the requested content within a given period, the user has to download additional segments from the network.

\subsection{Existing Studies}
A number studies have investigated caching placement optimization.
The existing studies can be categorized into two groups.

The investigations in  \cite{KPoularakis2016,YCui2016A,YCui2016Analysis,MXTaoContent2016,BZhou2016Stochastic,SMosleh2016,SYan2016User} considered caching at small cells. The works in \cite{YCui2016A,YCui2016Analysis,KPoularakis2016,MXTaoContent2016,BZhou2016Stochastic} jointly considered the caching and multicast technologies to optimize system performance. In \cite{KPoularakis2016}, the work investigated a multicast-aware caching problem.
The hardness of this problem was proved, and an algorithm with approximation ratio was proposed.
In \cite{YCui2016A}, the study developed a random caching design with multicasting in a large-scale cache-enabled wireless network.
An iterative algorithm was proposed to derive a local optimal solution. In order to reduce the computation complexity, an asymptotical optimal design was obtained.
Based on \cite{YCui2016A}, \cite{YCui2016Analysis} further investigated caching and multicast design with backhaul constraints in heterogeneous networks (HetNets).
In \cite{MXTaoContent2016}, the work considered a scenario with content-centric BS clustering and multicast beamforming.
The authors target optimizing the weighted sum of backhaul cost and transmit power.
In \cite{BZhou2016Stochastic}, a stochastic content multicast problem, originated from a Markov decision process, was formulated and a low-complexity algorithm was proposed. An assumption in \cite{MXTaoContent2016} and \cite{BZhou2016Stochastic} is
that the content placement was given. Relaxing the assumption, \cite{SMosleh2016} optimized the caching placement and proposed a mesh adaptive direct search algorithm.

Compared with caching at the small cells, the investigations in \cite{DMalak2016Optimizing,ZChen2016C,ZChen2016,HKang2014Mobile,JRao2016Optimal,Ji2016Wireless,Ji2016Fundamental} considered caching at the UEs, e.g., D2D caching networks. The studies in \cite{DMalak2016Optimizing,ZChen2016C,ZChen2016,HKang2014Mobile} analyzed and investigated caching problems by using stochastic geometry tools. In \cite{DMalak2016Optimizing}, the study investigated the optimal caching placements to maximize the average successful receptions' density.
In \cite{ZChen2016C}, the performance between caching at the small cells and UEs were analyzed and compared. Numerical results manifested that the performance varies by the user density and the content popularity distribution. In \cite{ZChen2016} and \cite{HKang2014Mobile}, the works investigated optimization problems with respect to probabilistic caching placement and average caching failure probability for each content.
In \cite{JRao2016Optimal}, the study addressed a two-tier caching network in which a subset of UEs and small cells have cache capability. In \cite{Ji2016Wireless} and \cite{Ji2016Fundamental}, the authors proposed an accurate simulation model taking into account a holistic system design and investigated information theoretic bounds for D2D caching networks, respectively.

Although the above studies focused on the cache placement design to optimize network performance, they neglected the impact of user mobility on caching performance. This issue was recognized in \cite{RWang2016Mobility}. In \cite{KPoularakis2013Ex} and \cite{KPoularakis2017}, the authors investigated the caching placement problem taking into account user mobility in HetNets, with the objective of minimizing the probability that the macrocell has to serve a request. The intractability of this problem was proved, and the problem is then reformulated using mixed integer programming (MIP). Moreover, the authors derived an upper bound for the objective function and proposed a distributed algorithm.
In \cite{TWei2014MPCS} and \cite{HLi2016Mobility}, the studies investigated a mobility
and popularity-based caching strategy (MPCS) and a seamless radio access network cache handover framework based on a
mobility prediction algorithm (MPA), respectively.
In \cite{YGuan2014Mo}, assuming that the trajectories of mobile users are known in advance, the authors investigated mobility-aware content caching and proposed an algorithm with approximation ratio.
In \cite{RWang2016}, the work optimized caching placement to maximize the data offloading ratio. A dynamic programming algorithm was proposed to obtain the optimal solution in small-scale scenarios. Since the algorithm complexity increases exponentially, the authors first proved that the objective function is a monotone submodular function, and then proposed a greedy algorithm which can achieve an $1/2$ approximation.

The investigations in \cite{KPoularakis2017} and \cite{RWang2016} are the most related works to our study. However, the system setup in \cite{KPoularakis2017} addresses caching at base stations, which is different from our study where we investigate caching at mobile users with mobility. In comparison to \cite{RWang2016}, our problem formulation takes into account the cost of downloading from network and guarantees that the collected segments are non-overlapping, along with giving a rigorous problem complexity analysis. In addition, our computational approach provides performance benchmarking of any sub-optimal algorithm for up to medium-size system scenarios.


\subsection{Our Contributions}
We investigate the cost-optimal caching problem with user mobility for D2D networks. Our objective is to optimize caching placement so as to minimize
the expected cost of obtaining files of interest by collecting file segments. The main contributions are summarized as follows.
First, a cost-optimal caching problem (COCP) is modelled, taking into account the impact of user mobility, cache size, and the total
number of encoded segments.
Accounting for this number is important in order to ensure no duplicates in the collected segments.
Second, the hardness of the problem is proved. To the best of our knowledge, this is the first mathematical
proof for the complexity of this
type of problems. The proof is based on a reduction from the 3-satisfiability (3-SAT) problem \cite{garey1979computers}. Moreover, for
problem-solving, due to the nonlinearity and high complexity of the objective function in COCP, a linear lower-bounding function is derived,
yielding an
approximation of COCP (ACOCP). The ACOCP approach brings two advantages. On one hand, it enables the global optimal solution by
using an off-the-shelf integer linear programming algorithm that can deliver solutions for small-scale and
medium-scale system scenarios effectively. Second, and more importantly, it serves the purpose of performance benchmarking of
any sub-optimal algorithm.
To be specific, by this approach, the lower bound of global optimum of COCP can be obtained.
We are hence able to gauge the deviation from optimum for any sub-optimal algorithm, whereas pure heuristics algorithm cannot be used for
such a purpose.
To tackle large-scale scenarios, it is proved that the optimal caching placement of one user, giving other users' caching placements,
can be derived in polynomial time.
Then, based on this proof, a mobility aware user-by-user (MAUU) algorithm is developed. Finally, Simulations are conducted
to verify the effectivenesses of the ACOCP approach and the MAUU algorithm by comparing them to the lower bound of global optimum and conventional
caching algorithms. Simulation results manifest that solving ACOCP leads to an effective approximation scheme -- the solution of ACOCP
does not deviate more than $4.4\%$ from the global optimum of COCP. The true performance figure is likely to be better because the performance evaluation
 is derived using the lower bounds. For the MAUU algorithm, the gap value to global optimum is less than $9\%$.
 Thus, the algorithm achieves excellent balance between complexity and accuracy.
In addition, the proposed algorithms also significantly outperform conventional caching algorithms, especially for large-scale scenarios.

The remainder of this paper is organized as follows. Section~\ref{System_model} introduces the system scenario, assumptions for caching placement,
and cost model. Section~\ref{NPHard} first derives the problem formulation, and then provides a rigorous complexity analysis.
Section~\ref{Simplified_formulation} presents the lower bound approximation approach of COCP.
Section~\ref{Heuristic_algorithm} develops an fast yet effective mobility aware user-by-user algorithm.
Performance evaluation is presented in Section~\ref{Performance_evaluation}. Finally, Section~\ref{Conclusions} concludes this paper.
\begin{figure*}[htbp]
\centering
\includegraphics[scale=0.6]{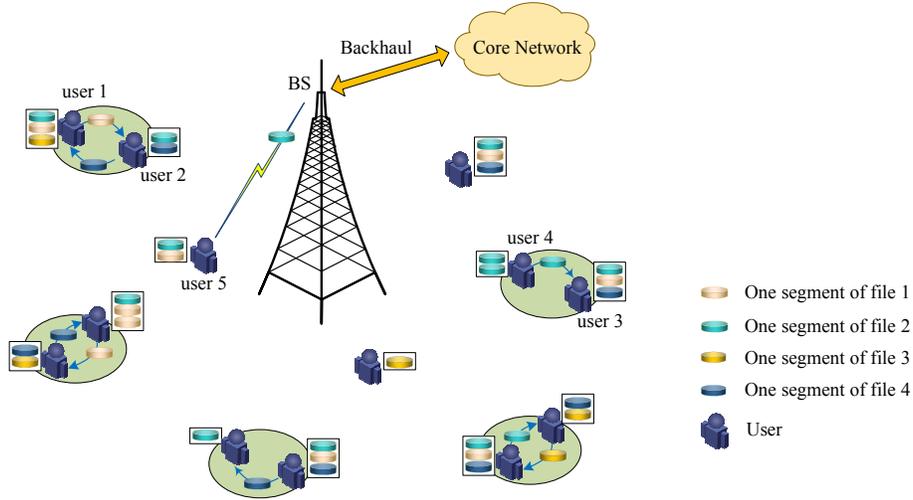}
\begin{center}
\caption{System scenario.}
\label{Rese_Scenario}
\end{center}
\end{figure*}
\section{System Model} \label{System_model}
\subsection{System Scenario}
 There are a total of $U$ mobile users in a network, whose index set is represented by $\mathcal{U}=\{1,2,\dots,U \}$.
 Each user $i$, $i \in \mathcal{U}$, is equipped with a cache of size $C_i$. Fig.~\ref{Rese_Scenario} shows the system scenario in which
 mobile users are able to collect the content when they meet each other, e.g., user 1 and user 2.

The inter-contact model has been widely used to describe the mobility pattern of mobile users \cite{VConan2008,APassarella2013}.
In this model, the mobile users can communicate with each other when they meet.
The contact process between any two mobile users is characterized by points along a timeline.
Each point represents a time that the two users meet, and the inter-contact time represents the time between two consecutive points.
The inter-contact time for any two users follows an exponential distribution. Moreover, it is assumed that the processes for the user pairs are independent.
Hereafter, the term contact is used to refer to the event that two users meet each other.
\subsection{Caching Placement}
There are a total of $F$ files, whose index set is represented by $\mathcal{F}=\{1,2,\dots,F \}$.
Each file $f$, $f\in \mathcal{F}$, is encoded into $S^f_\text{max}$ segments through a coding technique \cite{KPoularakis2017,Leong2012Distributed}. File $f$ can be recovered by collecting at least $S^f_\text{rec}$ distinct segments.
To describe the caching solution, we define a caching placement vector $\bm{x}$:
\[
\bm{x}= \{x_{fi}\in \mathbb{N}, f\in \mathcal{F}, i\in \mathcal{U} \},
\]
where $x_{fi}$ represents the number of segments of file $f$ stored at the user $i$.
Denote by $P_{fi}$ the probability that user $i$ requests file $f$, with $\sum_{f=1}^{F}P_{fi}=1$. When user $i$ requests file $f$, it will collect the segments of the file from its own cache and from the encountered users through D2D communications. The latter is subject to a time period $T_\text{D}$. For example, in Fig.~\ref{Rese_Scenario}, user 1 will collect one segment of file 4 from user 2. At the same time, user 2 will collect one segment of file 1 from user 1. But user 4 cannot collect the content of file 3 from user 3, because the latter does not store any segment of file 3.

Each user will check the total number of collected segments of the requested file at the end of $T_\text{D}$.
If the total number of collected segments of file $f$ is at least $S^f_\text{rec}$, user $i$ can recover this file. Otherwise, user $i$ will have to download additional segments from the network in order to reach $S^f_\text{rec}$ segments, e.g., user 5 in the figure.
The file recovering process considers only segments that are distinct from each other in the cache. For example, user 4 stores two distinct segments of file 2.

\subsection{Cost Model}
Up to $B$ segments can be collected by each user when two users meet.
Denote by $M_{ij}$ the number of contacts for users $i$ and $j$. Here, $M_{ij}$ follows a Poisson distribution with mean $\lambda_{ij}T_\text{D}$, where $\lambda_{ij}$ represents the average number of contacts per unit time. The number of segments of file $f$ collected by user $i$ from user $j$ within $T_\text{D}$, denoted by $S_{fij}$, is $\min(BM_{ij},x_{fj})$.
The number of segments of file $f$ collected by user $i$ from itself and all the other users via contacts within $T_\text{D}$, denoted by $S_{fi}$, is given as
\[
S_{fi}=\underset{j\in \mathcal{U},j\neq i}\sum \min(BM_{ij},x_{fj})+x_{fi}.
\]
If $S_{fi}<S^f_\text{rec}$, user $i$ will download $S^f_\text{rec}-S_{fi}$ segments from the network. This entity for file $f$ and user $i$, denoted by $S^N_{fi}$, is thus
$\max(S^f_\text{rec}-S_{fi},0)$.
Denote by $\delta_\text{D}$ and $\delta_\text{N}$ the costs of obtaining one segment from a user and the network, respectively. The cost for user $i$ to recover file $f$, denoted by $\Delta_{fi}$, is
$(S_{fi}-x_{fi})\delta_\text{D}+S^N_{fi}\delta_\text{N}$.
Taking into account the distribution of file request probabilities, the cost for user $i$ to recover its requested files, denoted by $\Delta_i$, is
$\underset{f\in \mathcal{F}}\sum P_{fi}\Delta_{fi}$.
Thus, the expected average cost per user can be expressed as
\begin{equation}
\begin{aligned}
 \Delta = &\text{E}\{\frac{1}{U}\underset{i\in \mathcal{U}}\sum \Delta_i \}\\
=& \text{E}\{\frac{1}{U}\underset{i\in \mathcal{U}}\sum \underset{f\in \mathcal{F}}\sum P_{fi}[ (S_{fi}-x_{fi})\delta_\text{D}+ \max(S^f_\text{rec}-S_{fi},0)\delta_\text{N} ] \}.
\label{C}
\nonumber
\end{aligned}
\end{equation}

\section{Problem Formulation and Complexity Analysis}\label{NPHard}
\subsection{Problem Formulation}
Our problem is to minimize $ \Delta$ by optimizing $\bm{x}$.
Thus, the cost-optimal caching problem (COCP) can be formulated as
\vskip 10pt
\begin{figure}[!h]
\vskip -20pt
\begin{subequations}
\begin{alignat}{2}
\quad &
\min\limits_{\bm{x}}\quad   \text{E}\{\frac{1}{U}\underset{i\in \mathcal{U}}\sum \underset{f\in \mathcal{F}}\sum P_{fi}[ (S_{fi}-x_{fi})\delta_\text{D}+ \max(S^f_\text{rec}-S_{fi},0)\delta_\text{N} ] \}
 \label{F_e} \\
\text{s.t}. \quad
& \underset{f\in \mathcal{F}}\sum x_{fi}\le C_i, ~i\in \mathcal{U} \label{F_f}\\
& \underset{i\in \mathcal{U}}\sum x_{fi}\le S^f_\text{max}, ~f\in \mathcal{F} \label{F_h}\\
& x_{fi} \in \mathbb{N}, ~i\in \mathcal{U}, ~ f\in \mathcal{F} \label{F_g}
\end{alignat}
\label{eq:mine}
\vskip -20pt
\end{subequations}
\end{figure}

Eq.~(\ref{F_f}) requires the total number of cached segments to adhere to cache capacity limit. By Eq.~(\ref{F_h}), the total number of segments of a file, cached by all users, does not exceed the number of encoded segments.
This constraint guarantees that the collected segments of any file will be distinct from each other.

\subsection{Complexity Analysis}
\noindent \textbf{Theorem 1.} \textbf{COCP} is $\mathcal{NP}$-hard.

\emph{Proof:} We adopt a polynomial-time reduction from the 3-satisfiability (3-SAT) problem that is $\mathcal{NP}$-complete. Consider any 3-SAT instance with $m$ Boolean variables $z_1, z_2, \dots, z_m$, and $n$ clauses. A variable or its negation is called a literal. Denote by $\hat{z}_i$ the negation of $z_i$, $i=1,2, \dots, m$. Each clause consists of a disjunction of exactly three different literals, e.g., $\hat{z}_1 \lor z_2 \lor z_3$. The 3-SAT problem amounts to determining whether or not there exists an assignment of true/false values to the variables, such that all clauses are satisfied (i.e., at least one literal has value true in every clause). It is assumed that no clause contains both a variable and its negation; such clauses become always satisfied, thus they can be eliminated by preprocessing. Moreover, a literal appears in at least one clause as otherwise the corresponding value assignment is trivial. For the same reason, a literal is present in at most $n-1$ clauses.

We construct a reduction from the 3-SAT instance as follows. The number of users is $U=2m+n$, referred to as literal and clause users, respectively, i.e., $\mathcal{U}=\{1, 2, \dots, 2m+n\}$. There are two files $a$ and $b$, i.e., $\mathcal{F}=\{a, b \}$, each of them has $m$ segments, i.e., $S^a_\text{max}=S^b_\text{max}=m$. File $a$ or $b$ can be recovered by collecting one segment, i.e., $S^a_\text{rec}=S^b_\text{rec}=1$. The cache size of literal and clause users are one ($C_i=1$, $i=1,2,\dots,2m$) and zero ($C_j=0$, $j=2m+1,\dots,2m+n$), respectively.

The literal users are formed into $m$ pairs. Denote by $\epsilon$ a small positive number. We set $\delta_\text{N}>3n+\frac{n\epsilon(m-3)}{(1-\epsilon)^{m-2}}\delta_\text{D}$, and
$\lambda_{ij}=\ln(\frac{1}{\epsilon})$ for users $i$ and $j$ in each of the $m$ pairs. Then these users meet at least once with probability $1-\epsilon$. We set $\lambda_{ij}=\ln{\frac{1}{1-\epsilon}}$ for the other literal users where $i$ and $j$ are from different pairs, so that these users meet at least once with probability $\epsilon$. Each literal user is interested in downloading both files $a$ and $b$ with equal probability, i.e., $P_{ai}=P_{bi}=1/2$, $i=1,\dots,2m$.
First, suppose one of the users in each pair caches file $a$, and the other caches file $b$, or vice versa. It means that for any pair, the caching content is either $ab$ or $ba$. This corresponds to the Boolean value assignment in the original 3-SAT instance. In such a case, the expected cost that both users of a pair recover both files $a$ and $b$, denoted by $ \Delta_1$, is given as
\[
 \Delta_1=(1-\epsilon)\delta_\text{D}+2\epsilon(m-1)\delta_\text{D}+\epsilon(1-\epsilon)^{m-1}\delta_{N}.
\]
Consequently, the total cost for all the literal users, denoted by $ \Delta^l_1$, is $m \Delta_1$.

Each clause user is interested in downloading file $a$ with probability one, i.e., $P_{ai}=1$, $i=2m+1,\dots,2m+n$. If users $i$ and $j$ are clause users, $\lambda_{ij}$ can be anything as their all have a cache size of zero. For a clause user $i$, if $j$ is one of the three literal users in the corresponding clause in the 3-SAT instance, we set $\lambda_{ij}=\ln(\frac{1}{\epsilon})$. Otherwise, we set $\lambda_{ij}=\ln(\frac{1}{1-\epsilon})$.
If at least one of the three literal users caches file $a$, then the expected cost for a clause user is at most
$3(1-\epsilon)\delta_\text{D}+\epsilon(m-3)\delta_\text{D}+\epsilon^3(1-\epsilon)^{m-3}\delta_\text{N}$.
The corresponding values for the $n$ clause users together, denoted by $ \Delta^c_1$, is $n(3(1-\epsilon)\delta_\text{D}+\epsilon(m-3)\delta_\text{D}+\epsilon^3(1-\epsilon)^{m-3}\delta_\text{N})$.

By the construction above, which is polynomial, the cost is no more than $ \Delta^l_1+ \Delta^c_1$ if the 3-SAT instance is satisfiable. Otherwise, at least one clause user has virtually no other option, than downloading from the network and the expected total cost is at least $m \Delta_1+(n-1) \Delta{'}+\epsilon(m-3)\delta_\text{D}+(1-\epsilon)^{m-3}\delta_\text{N}> \Delta^l_1+ \Delta^c_1$, where $ \Delta{'}=(1-\epsilon)\delta_\text{D}+\epsilon(m-3)\delta_\text{D}+\epsilon(1-\epsilon)^{m-3}\delta_\text{N}$.
Thus, whether or not there exists a caching placement strategy with a total expected cost of no more than $ \Delta^l_1+ \Delta^c_1$ gives the correct answer to 3-SAT.

Now, let's consider the case where some of the literal user pairs cache the same file.
If there is one pair caching file $a$, i.e., the caching content is $aa$, another pair cache $bb$, and the remaining pairs cache either $ab$ or $ba$. The total literal users' cost, denoted by $ \Delta^l_2$, is given as
\[
\Delta^l_2=2((1-\epsilon)\delta_\text{D}+2\epsilon \delta_\text{D}+2(m-2)\epsilon \delta_\text{D}+(1-\epsilon)^m \delta_\text{N})+(m-2) \Delta_1.
\]
If all the clause users can obtain file $a$ from the literal users, the total clause users cost, denoted by $ \Delta^c_2$, is no less than $n \Delta{'}$.
The corresponding values for all the users together is $ \Delta^l_2+ \Delta^c_2$, and $ \Delta^l_2+ \Delta^c_2> \Delta^l_1+\Delta^c_1$. If there is more than one pair caching the same file, e.g., two pairs cache $aa$, the cost becomes even higher.
Thus, the previous conclusion remains valid, namely whether or not there is an assignment with no more than $ \Delta^l_1+ \Delta^c_1$ gives the right answer even this case included.

Therefore, the recognition versions of COCP is $\mathcal{NP}$-complete and its optimization version is $\mathcal{NP}$-hard. 
$\hfill{} \Box$


\section{Lower Bound Approximation Approach} \label{Simplified_formulation}
Due to the COCP's intractability, generally it is difficult to obtain the global optimal solution. For problem-solving, we linearize the
first part of objective function and derive a lower bound for the second part. These together give us a linear lower-bounding function,
as an approximation to the original function. As a result, the problem can be reformulated as a mixed linear integer program.

Define
\begin{equation}
\begin{aligned}
 \Delta^{lb} \triangleq \frac{1}{U}\underset{i\in \mathcal{U}}\sum \underset{f\in \mathcal{F}}\sum P_{fi}[ \Delta^\text{d}_{fi}+\max( \Delta^\text{n}_{fi},0)],
\label{Linear}
\end{aligned}
\end{equation}
and
\begin{equation}
\left\{
\begin{aligned}
{}&\Delta^\text{d}_{fi}= \text{E}(\underset{j\in \mathcal{U},j\neq i}\sum \min(BM_{ij},x_{fj}))\delta_\text{D},\\
{}&\Delta^\text{n}_{fi}=S^f_\text{rec}\delta_\text{N}-\text{E}[\underset{j\in \mathcal{U},j\neq i}\sum \min(BM_{ij},x_{fj}) +x_{fi}]\delta_\text{N}.
\label{XX3}
\end{aligned}
\right.
\end{equation}

\noindent \textbf{Theorem 2.} $\Delta^{lb}$ is a lower-bounding function of $\Delta$, i.e.,
\[
 \Delta \ge  \Delta^{lb}.
 \label{C1C2}
\]

\emph{~~~~Proof:} See Appendix \ref{Appex_Lower}. $\hfill{} \Box$

Using $\Delta^{lb}$, an approximation of COCP (ACOCP) can be formulated as
\vskip 10pt
\begin{figure}[!h]
\vskip -20pt
\begin{subequations}
\begin{alignat}{2}
\quad &
\min\limits_{\bm{x}}\quad  \frac{1}{U}\underset{i\in \mathcal{U}}\sum \underset{f\in \mathcal{F}}\sum P_{fi}[\Delta^\text{d}_{fi}+\max(\Delta^\text{n}_{fi},0)] \\
\text{s.t}. \quad
\label{F_e1}
& \underset{f\in \mathcal{F}}\sum x_{fi}\le C_i, ~i\in \mathcal{U}\\
& \underset{i\in \mathcal{U}}\sum x_{fi}\le S^f_\text{max}, ~f\in \mathcal{F}\\
& x_{fi} \in \mathbb{N}, ~i\in \mathcal{U}, ~ f\in \mathcal{F}
\end{alignat}
\label{eq:mine}
\vskip -20pt
\end{subequations}
\end{figure}

To obtain the above problem's global optimal solution, we introduce binary variable $y^k_{fi}$ that is one if and only if user $i$
caches $k$ segments of file $f$.
Denote by $\bm{y}$ the vector consisting of $y^k_{fi}$:
\[
\bm{y}=\{y^k_{fi}, ~i\in \mathcal{U}, ~ f\in \mathcal{F},~k\in [0,S^f_\text{rec}]\}.
\]
By definition, if $x_{fi}=k$, then $y^k_{fi}=1$. For example, if $x_{fi}=3$, then $y^3_{fi}=1$ and $y^k_{fi}=0$ for the case that
$k\not=3$.
Thus, the relationship between the optimization variables $x_{fi}$ and $y^k_{fi}$ can be expressed as
\begin{equation}
\left\{
\begin{aligned}
{}& x_{fi}=\sum_{k=0}^{S^f_\text{rec}}ky^k_{fi},~i\in \mathcal{U}, ~ f\in \mathcal{F},\\
{}& \sum_{k=0}^{S^f_\text{rec}}y^k_{fi}=1,~i\in \mathcal{U}, ~ f\in \mathcal{F}.
\label{xy}
\end{aligned}
\right.
\end{equation}
Define
\begin{equation}
\begin{aligned}
e^k_{fij}\triangleq &\text{E}(\min(BM_{ij},k))\\
=&\sum_{t=0}^{k}t\text{Pr}(BM_{ij}=t)+k\text{Pr}(BM_{ij}>k),
\end{aligned}
\end{equation}
where
\begin{equation}
\text{Pr}(BM_{ij}=t)=\left\{
\begin{aligned}
{}& \frac{(\lambda_{ij}T_\text{D})^{\frac{t}{B}}e^{-\lambda_{ij}T_\text{D}}}{\frac{t}{B}}, ~\text{if}~(t~\text{mod}~B)=0,\\
{}& 0,~\text{else}.
\label{Mij}
\end{aligned}
\right.
\end{equation}
Thus, for any $x_{fj}$, $\text{E}(\min(BM_{ij},x_{fj}))$ can be expressed as
\[
\text{E}(\min(BM_{ij},x_{fj}))=\sum_{k=0}^{S^f_\text{rec}} e^k_{fij}y^k_{fj}.
\]
Moreover, by the proof in Appendix \ref{Appex_Lower}, it follows that
\[
\Delta^\text{n2}_{fi}=\max(\Delta^\text{n}_{fi},0).
\]
Therefore, through the above mathematical analysis, ACOCP can be reformulated as
\vskip 10pt
\begin{figure}[!h]
\vskip -20pt
\begin{subequations}
\begin{alignat}{2}
\quad &
\min\limits_{\bm{y}}\quad  \frac{1}{U}\underset{i\in \mathcal{U}}\sum \underset{f\in \mathcal{F}}\sum P_{fi}(\Delta^\text{d}_{fi}+\Delta^\text{n2}_{fi}) \\
\text{s.t}. \quad
\label{F_b}
&\Delta^\text{n2}_{fi} \ge \Delta^\text{n}_{fi},~ i\in \mathcal{U}, ~ f\in \mathcal{F}\\
\label{F_c}
&\Delta^\text{n2}_{fi} \ge 0, ~ i\in \mathcal{U},~ f\in \mathcal{F}\\
\label{F_e}
& \sum_{k=0}^{S^f_\text{rec}}y^k_{fi}=1,~i\in \mathcal{U}, ~ f\in \mathcal{F}\\
& \underset{f\in \mathcal{F}}\sum \sum_{k=0}^{S^f_\text{rec}}ky^k_{fi}\le C_i, ~i\in \mathcal{U}\\
& \underset{i\in \mathcal{U}}\sum \sum_{k=0}^{S^f_\text{rec}} ky^k_{fi}\le S^f_\text{max}, ~f\in \mathcal{F}\\
& y^k_{fi} \in \{0,1 \},~i\in \mathcal{U}, ~ f\in \mathcal{F},~k\in [0,S^f_\text{rec}]
\end{alignat}
\label{eq:mine}
\vskip -20pt
\end{subequations}
\end{figure}

where
\begin{equation}
\left\{
\begin{aligned}
{}&\Delta^\text{d}_{fi}=\underset{j\in \mathcal{U},j\neq i}\sum \sum_{k=0}^{S^f_\text{rec}} (e^k_{fij}y^k_{fj})\delta_\text{D},\\
{}&\Delta^\text{n}_{fi}=S^f_\text{rec}\delta_\text{N}-\underset{j\in \mathcal{U},j\neq i}\sum \sum_{k=0}^{S^f_\text{rec}} (e^k_{fij}y^k_{fj}) \delta_\text{N}-\sum_{k=0}^{S^f_\text{rec}} (ky^k_{fi})\delta_\text{N}.
\label{X3}
\end{aligned}
\right.
\end{equation}
Note that the definitions of $\Delta^\text{d}_{fi}$ and $\Delta^n_{fi}$ are the reformulations of that in (\ref{XX3}).

The above objective function and constraints are linear with respect to $\bm{y}$. Thus, the ACOCP approach can use an off-the-shelf integer programming
algorithm
from optimization packages, e.g., Gurobi \cite{Gurobi}, to obtain the global optimal solution. Generally, it can deliver
optimal solutions for the small-scale and medium-scale system scenarios effectively. What's more, it serves the purpose
of performance benchmarking of any sub-optimal algorithm. Denote by $\bm{y}^*$ the global optimal solution of ACOCP.
By (\ref{xy}), $\bm{y}^*$ can be converted into an approximation solution of COCP, referred to as $\bm{x}^{lb}$. Denote by
$\bm{x}^*$ the global optimal solution of COCP. By Theorem 2, it follows that
\begin{equation}
\left\{
\begin{aligned}
& \Delta(\bm{x}^{lb})\ge  \Delta(\bm{x}^*),\\
& \Delta(\bm{x}^*)\ge \Delta^{lb}(\bm{x}^*),\\
& \Delta^{lb}(\bm{x}^*)\ge  \Delta^{lb}(\bm{x}^{lb}).
\end{aligned}
\right.
\end{equation}
Therefore,
\begin{equation}
\begin{aligned}
 \Delta(\bm{x}^{lb})\ge  \Delta(\bm{x}^*)\ge  \Delta^{lb}(\bm{x}^{lb}).
\label{UpDown}
\end{aligned}
\end{equation}
 Eq. (\ref{UpDown}) indicates that if $\bm{x}^{lb}$ is derived, a lower bound, $\Delta^{lb}(\bm{x}^{lb})$, of global optimum of COCP
 is obtained.
 The lower bound can be used to evaluate the optimality deviation of the solution of  ACOCP. Namely, the gap between the approximation solution and
 the global optimal solution of COCP does not exceed $ \Delta(\bm{x}^{lb})-  \Delta^{lb}(\bm{x}^{lb})$, while heuristic algorithms cannot
 provide this type of performance assessment. More importantly, it can evaluate the solution of any sub-optimal algorithm, such as the
 one presented in the next section, because the gap to the global optimum does not exceed the gap to the lower bound.



\section{Mobility Aware User-by-User Algorithm} \label{Heuristic_algorithm}
Although the ACOCP approach can obtain solutions for up to medium-size scenarios, the computation complexity does not scale well.
Thus, we propose a fast yet effective algorithm, i.e., mobility aware user-by-user (MAUU) algorithm. A general description of MAUU is as follows. The users are treated
one by one starting with the first user. Initially, the caching content of all the users are set to be empty.
The algorithm optimizes the caching content of the first user, and then keeps this content fixed for this user in later iterations while
performing the optimization for the other users.
Once the cache content allocation of one user is optimized, the remaining segments of each file, denoted by $S^f_\text{rem}$, $f\in \mathcal{F}$,
will be updated accordingly. The same process repeats for the next user.

 \begin{algorithm}
\caption{The MAUU algorithm for \textbf{COCP}}
\label{alg1}
\begin{algorithmic}[1]
\REQUIRE $\mathbf{S}_\text{rem}$, $\mathbf{S}_\text{rec}$, $\bm{x}$, $\bm{x}_1$, $\mathbf{C}$, $U$, $F$, $B$, $\delta_\text{D}$, $\delta_\text{N}$.
\ENSURE $\bm{x}$

\FOR{$i=1$ : $U$}
\STATE $\bm{g} \leftarrow  \emptyset$, $\mathbf{V} \leftarrow  [0]_{F\times C_i}$, and $\mathbf{W} \leftarrow  [0]_{C_i \times F}$
 \FOR{$f=1$ : $F$}
   \FOR{$k=0$ : $\min(C_i, \mathbf{S}_\text{rec}(f), \mathbf{S}_\text{rem}(f))$}
\STATE $x_1(f,i) \leftarrow k$
\STATE $v(f,k) \leftarrow  \Delta(\bm{x}_1)$
\STATE $x_1(f,i) \leftarrow 0$
\ENDFOR
\ENDFOR

\FOR{$q=1$ : $F$}

\IF{$q <F$}

\FOR{$k{'}=0$ : $C_i$}

\IF {$q=1$}

\STATE $w(1,k{'}) \leftarrow v(1,\min(k{'},\mathbf{S}_\text{rec}(1),\mathbf{S}_\text{rem}(1))) $
\STATE $\bm{g}_{1k{'}} \leftarrow \{\min(k{'},\mathbf{S}_\text{rec}(1),\mathbf{S}_\text{rem}(1))\}$

\ELSE
\STATE $w(q,k{'}) \leftarrow \argmin\{v(q,r_q)+w(q-1,k{'}-r_q), r_q=0,1,\dots,\min(k{'}, \mathbf{S}_\text{rec}(q), \mathbf{S}_\text{rem}(q))\}$
\STATE $\bm{g}_{qk{'}} \leftarrow \bm{g}_{q-1,k{'}-r^*_q} \cup \{r^*_q  \}$
\ENDIF

\ENDFOR

\ELSE

\STATE $w(F,C_i) \leftarrow \argmin\{v(F,r_F)+w(F-1,C_i-r_F), ~~~r_F=0,1,\dots,\min(C_i, \mathbf{S}_\text{rec}(F), \mathbf{S}_\text{rem}(F))\}$
\STATE $\bm{g}_{FC_i} \leftarrow \bm{g}_{F-1,C_i-r^*_F} \cup \{r^*_F  \}$

\ENDIF

\ENDFOR
 \STATE$\mathbf{S}_\text{rem} \leftarrow \mathbf{S}_\text{rem}-\bm{g}_{FC_i}$
 \STATE $\bm{x}^i \leftarrow \bm{g}_{FC_i}$ \\
 \STATE$\bm{x}_1 \leftarrow  \bm{x}$\\
\ENDFOR

\RETURN $\bm{x}$

\end{algorithmic}
\end{algorithm}
\subsection{Optimal Caching for One User}
\noindent \textbf{Theorem 3.} Optimizing the caching placement of one user can be derived in polynomial time when the caching placements
 of the other users are given.

\emph{Proof:} We compute a matrix, called cost matrix and denoted by $\mathbf{V}$, in which entry $v(f,k)$ represents the current expected
total cost if this user caches $k$ segments of file $f$.
The entries of this matrix can be computed using Eq. (\ref{SimObjeFun}) in Appendix B.

Below a recursive function is introduced to derive the optimal caching placement for the user. We define a second  matrix, called the
optimal cost matrix, and denote it by $\mathbf{W}$, in which $w(q,k{'})$ represents the cost of the optimal solution from considering
the first $q$ files using a cache size of $k{'}$, $k{'}=0,1,\dots,C$; here $C$ denotes the cache size of the user under consideration.
The value of $w(q,k{'})$ is given by the following recursion:
 \begin{equation}
\begin{aligned}
w(q,k{'})=\underset{r}{\argmin}\{ v(q,r)+w(q-1,k{'}-r)\},
\label{recurive equation}
\end{aligned}
\end{equation}
where $r$ can vary from $0$ to at most $\min\{k{'}, S^q_\text{rec}, S^q_\text{rem}\}$ due to cache size $k{'}$, file recovery
threshold $S^q_\text{rec}$, and the number of remaining segments $S^q_\text{rem}$ of file $q$.
Using Eq.~(\ref{recurive equation}), the optimal cost for file $q$ is computed when the optimal cost of the first $q-1$ files
is given.

For the overall solution, the optimal cost can be computed using the above recursion for cache size of $C$ and $F$ files. We prove it
by mathematical induction.
First, when $q=1$, obviously $w(1,k{'})=\underset{r}{\argmin}\{ v(1,r)\}$ for all $k{'}$. There are $\min\{k{'}, S^q_\text{rec}, S^q_\text{rem}\}+1$
possible values of $r$, and considering these values one by one gives the optimum $r^*$.
Now, assume $w(l,k{'})$ is optimal for some $l$. We prove that $w(l+1,k{'})$ is optimal. According to the recursive function,
\[
w(l+1,k{'})=\underset{r}{\argmin}\{ v(l+1,r)+w(l,k{'}-r)\}.
\]
The possible values for $r$ is from $0$ to $\min\{k{'}, S^q_\text{rec}, S^q_\text{rem}\}$, and for each of the possible values of $r$, $w(l,k{'}-r)$ is optimal. This together gives the conclusion that the minimum will be obtained indeed by the $\argmin$ operation.
Thus, $w(q,k{'})$ is optimal.

Finally, we show that $w(F,C)$ can be computed in polynomial time.
By Appendix B, the complexity of computing $\mathbf{V}$ is of $O(CF^2U^2S{'}^2_\text{rec})$.
By the above, the computational complexity of $\mathbf{W}$ is of $O(FC^2)$.
Thus, optimizing the cache content of one user runs in $O(CF^2U^2S{'}^2_\text{rec})+O(FC^2)=O(CF^2U^2S{'}^2_\text{rec})$ because generally $FU^2S{'}^2_\text{rec}>C$.
$\hfill{} \Box$
\subsection{Algorithm Summary}
The algorithmic flow is presented in Algorithm 1. The input parameters consist of
$\mathbf{S}_\text{rem}$, $\mathbf{S}_\text{rec}$, $\bm{x}$, $\bm{x}_1$, $\mathbf{C}$, $U$, $F$, $B$, $\delta_\text{D}$,
and $\delta_\text{N}$.
Here, $\mathbf{S}_\text{rem}$ is a vector consisting of the remaining segments of all the files. The initialization step is to set
$\mathbf{S}_\text{rem}=\{S^1_\text{max},\dots,S^F_\text{max}\}$, $\mathbf{S}_\text{rec}=\{S^1_\text{rec},\dots,S^F_\text{rec}\}$,
and $\mathbf{C}=\{C_1,\dots,C_U \}$.
The final caching placement solution is again denoted by $\bm{x}$. However, for the convenience of description, our algorithm treats it
 as a matrix of size $F\times U$. We also define $\bm{x}_1$ as an auxiliary matrix with the same size as $\bm{x}$.
Initially, the algorithm sets all the entries of $\bm{x}$ and $\bm{x}^{\prime}$ to zero, i.e., $\bm{x}=[0]_{F\times U}$ and
$\bm{x}_1=[0]_{F\times U}$.

For a generic iteration for one user, denote by $r_{q}^*$ the optimal number of segments cached for file $q$, and denote by vector
 $\bm{g}_{qk{'}}$ the optimal caching placement for the user under consideration with cache size of $k{'}$ and first $q$ files.
By Line 1, the users are processed one by one.
Line 2 initializes $\mathbf{V}$, $\mathbf{W}$, and $\bm{g}$.
Lines 3-7 compute matrix $\mathbf{V}$. Lines 8-19 compute $\mathbf{W}$ and $\bm{g}$. Lines 20-22 update $\mathbf{S}_\text{rem}$,
the $i$th column of $\bm{x}$ denoted by $\bm{x}^i$, and $\bm{x}_1$, respectively.



\section{Performance Evaluations} \label{Performance_evaluation}
We have developed two approaches that lead to solutions of COCP, i.e., the ACOCP approach and the MAUU algorithm. Next, simulations are conducted to evaluate the effectivenesses of the two approaches by comparing them to the lower bound of global optimum and conventional caching
 algorithms, i.e., random caching \cite{Balaszczyszy2015Optimal} and popular caching \cite{Ahlehagh2014Video}. The two conventional
 algorithms consider users one by one.
In the former, each user will cache files randomly with respect to the files' request probabilities.
That is, the higher the request probability of a file is, the more likely this file will be cached. In the latter, each user
will cache the files according to the popularity in terms of the files' request probabilities of this user. Besides, in implementing the two
algorithms, to ensure that the collected segments are non-overlapping,
the total number of cached segments of each file, for all the users together, does not exceed the number of available segments.


The file request probability follows a Zipf distribution \cite{ZChen2016,KPoularakis2013Ex}, i.e., $P_{fi}=\frac{f^{-\gamma_i}}{\underset{k\in \mathcal{F}}{\sum}k^{-\gamma_i}}$, where $\gamma_i$ is the Zipf parameter for user $i$. The number of segments for recovering a file $f$, $S^f_\text{rec}$, is randomly selected in $[1,S^*]$, where $S^*$ will vary in the simulations, and each file has the same  $\alpha=S^f_\text{max}/S^f_\text{rec}$. The average number of contacts per unit time for users $i$ and $j$, $i\not=j$, $\lambda_{ij}$, is generated according to a Gamma distribution $\Gamma(4.43,1/1088)$ \cite{APassarella2013}. In the simulations, $\gamma_i$ and $C_i$ are uniform, namely, $\gamma_i=\gamma$ and $C_i=C$ for all $i$.

\subsection{Performance Comparison}
\begin{figure}
\centering
\includegraphics[scale=0.5]{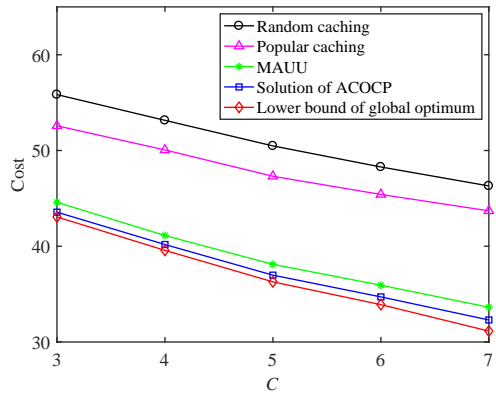}
\begin{center}
\caption{Impact of $C$ on $ \Delta $ when $U=8$, $F=80$, $B=1$, $\delta_\text{D}=1$, $\delta_\text{N}=30$, $\gamma=0.8$, $S^*=4$, $\alpha=3$, and $T_\text{D}=600s$.}
\label{Cchanged}
\end{center}
\end{figure}

\begin{figure}
\centering
\includegraphics[scale=0.5]{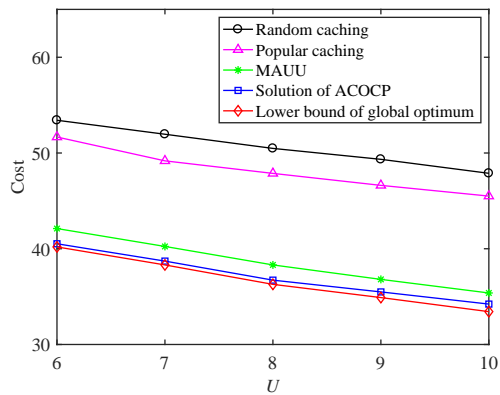}
\begin{center}
\caption{Impact of $U$ on $ \Delta $ when $F=80$, $C=5$, $B=1$, $\delta_\text{D}=1$, $\delta_\text{N}=30$, $\gamma=0.8$, $S^*=4$, $\alpha=3$, and $T_\text{D}=600s$.}
\label{userchanged}
\end{center}
\end{figure}

\begin{figure}
\centering
\includegraphics[scale=0.5]{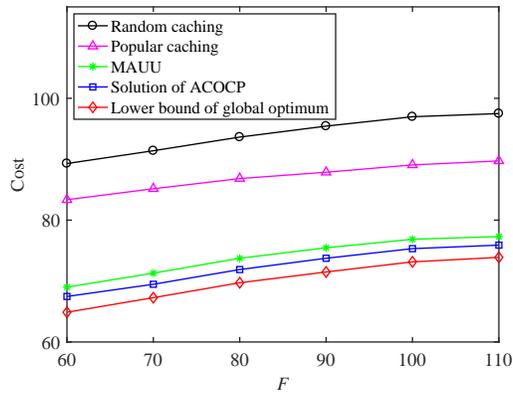}
\begin{center}
\caption{Impact of $F$ on $ \Delta $ when $U=8$, $B=2$, $\delta_\text{D}=1$, $\delta_\text{N}=30$, $C=5$, $\gamma=0.8$, $\alpha=3$, and $T_\text{D}=600s$.}
\label{Fchanged}
\end{center}
\end{figure}

\begin{figure}
\centering
\includegraphics[scale=0.5]{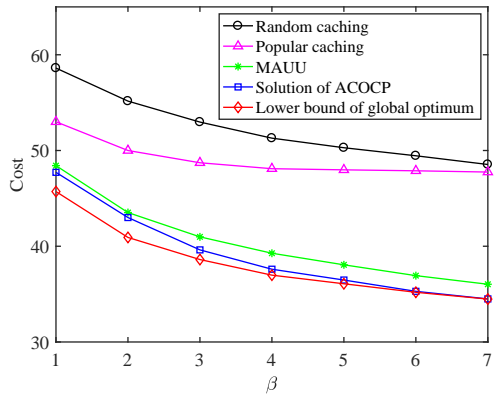}
\begin{center}
\caption{Impact of user average speed on $ \Delta $ when $U=8$, $F=80$, $B=1$, $\delta_\text{D}=1$, $\delta_\text{N}=30$, $C=5$, $S^*=4$, $\alpha=3$, $\gamma=0.8$, $\theta=1/1088$, and $T_\text{D}=600s$.}
\label{Betachanged}
\end{center}
\end{figure}

\begin{figure}
\centering
\includegraphics[scale=0.5]{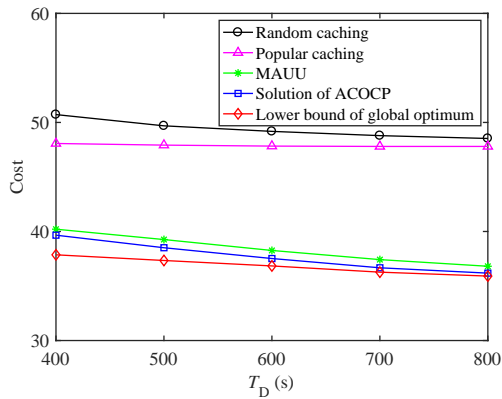}
\begin{center}
\caption{Impact of $T_\text{D}$ on $ \Delta $ when $U=8$, $F=80$, $C=5$, $B=2$, $\delta_\text{D}=1$, $\delta_\text{N}=30$, $\gamma=0.8$, $S^*=4$, and $\alpha=3$.}
\label{Tchanged}
\end{center}
\end{figure}

The performance of the ACOCP and the MAUU are shown in Figs.~\ref{Cchanged}-\ref{Tchanged}. The line in green and the line in blue denote the costs by using the MAUU algorithm and the solution of ACOCP (i.e., $\bm{x}^{lb}$), respectively. The line in red represents the cost of the lower bound of global optimum, i.e., $\Delta^{lb}(\bm{x}^{lb})$ in (\ref{UpDown}).

In general, the true optimality gaps of ACOCP approach and MAUU algorithm (or any sub-optimal algorithm) are hard to get, because it is difficult to know the value of global optimum. However, by Section IV, the ACOCP approach provides an effective bound for performance evaluation, because the gap to the global optimum does not exceed the gap to the lower bound.

Fig.~\ref{Cchanged} and Fig.~\ref{userchanged} show the impact of $C$ and $U$, respectively. Overall, the cost linearly
decreases with respect to $C$ and $U$. This is expected, because the users can store more contents with the increase of cache size, and they have
more choices and consequently more possibility to collect the needed segments when the number of users grows.
In addition, when $C$ and $U$ increase, for the ACOCP approach, the solution is close to the lower bound of global optimum overall, but the gap to the bound increases slightly. For example, by increasing $C$ from $3$ to $7$, the gap grows from $0.91\%$ to $2.83\%$.
The reason is that, although the global optimal solution of ACOCP can be derived, it is a sub-optimal solution for COCP.
Increasing $C$ and $U$ leads to larger solution space and may make the bound weaker. However, the worsening is not significant.
Similarly, for the MAUU algorithm, the gap increases with the increase of the two parameters. This is because that, giving other users' caching placements,
the MAUU algorithm achieves the optimal solution of one user under consideration, whereas this solution is sub-optimal for the system. However, although increasing $C$ and $U$ may slightly decrease the accuracy, the MAUU algorithm remains promising, as the gap is lower than $9\%$.
Finally, the MAUU algorithm outperforms the conventional algorithms
consistently in the two figures, especially for big $U$ and $C$. When $C=7$, it outperforms the popular caching algorithm by $23.5\%$, and outperforms the random caching algorithm by $27.8\%$. Note that $U$ and $C$ represent the system size. Thus, the MAUU algorithm is useful for large-scale system scenarios.

The effect of $F$ is analyzed in Fig.~\ref{Fchanged}.
This figure shows results for which the number of segments for recovering a file $f$ is uniform, namely, $S^f_\text{rec}=4$ for any $f$.
It can be observed that the cost first grows with the increase of $F$. If $F$ becomes excessively big, the impact becomes insignificant due to the limit of cache size and the number of users. Besides,
when $F$ increases, the performance difference between the solution of ACOCP and the solution by using MAUU is fairly constant, but the popular caching algorithm outperforms the random caching algorithm significantly. Obviously,
increasing $F$ directly leads to higher diversity of files. Thus, for the random caching algorithm, the users are more likely to store the infrequently requested files.

The user average contact rate is proportional to the user average speed \cite{RWang2016}.
Thus, examining the impact of the former reflects also that of the latter. We generate the contact rate
for users $i$ and $j$, $\lambda_{ij}$, $i\not=j$, according to a Gamma distribution $\Gamma(\beta,\theta)$. Thus, the average contact
rate is $\beta\theta$.
Fig.~\ref{Betachanged} fixs $\theta$, and analyzes the impact of $\beta$ on $\Delta$.
A large average contact rate means more frequent contacts among users, resulted from high mobility.
The impact of $T_\text{D}$ is shown in Fig.~\ref{Tchanged}.
A greater $T_\text{D}$ indicates that the users have more time to collect the needed segments.
There are two common insights for the two figures.
First, the MAUU algorithm outperforms the popular caching algorithm.
When the values of the two parameters increase, the improvement is significant.
This is because the caching placement by MAUU can be dynamically adapted to the variations in the parameters, whereas the caching placement by popular caching is fixed.
Moreover, for the ACOCP approach, the gap to the lower bound progressively decreases. In particular, when $\beta=1$, the gap is $4.39\%$. While $\beta=6$, the gap decreases to $0.28\%$, indicating that the solution of ACOCP is very close to optimum. The reason is that in such cases, $\Delta^{lb}$ approaches $\Delta$.
\begin{figure}
\centering
\includegraphics[scale=0.5]{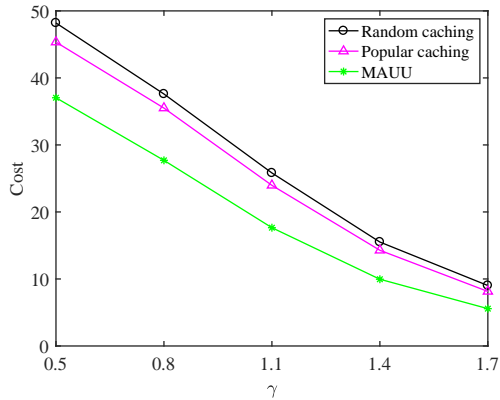}
\begin{center}
\caption{Impact of $\gamma$ on $ \Delta $ when $U=20$, $F=200$, $B=1$, $\delta_\text{D}=1$, $\delta_\text{N}=30$, $C=4$, $S^*=3$, $\alpha=3$, and $T_\text{D}=600s$.}
\label{Gammachanged}
\end{center}
\end{figure}
\subsection{Algorithm Scalability}
In Fig.~\ref{Gammachanged}, we show additional results for large-scale scenarios via increasing the number of users and files. Specifically, $U=20$ and $F=200$. For this case, we compare our scalable MAUU algorithm with the two conventional algorithms.
Overall, the costs of the caching solution from the MAUU algorithm and conventional caching algorithms exhibit the same decreasing trend with respect to $\gamma$.
The MAUU algorithm outperforms the two conventional caching algorithms as the latter algorithms neglect the effect
of user mobility.
However, there is an additional insight that the improvement of MAUU becomes smaller by increasing $\gamma$.
The reason is that for high $\gamma$, the files' request probability has a
large variation. As a result, the users are more inclined to request the popular files.

A general observation is that the ACOCP approach is more accurate than the MAUU algorithm -- the cost by using the solution of ACOCP is always less than that of MAUU.
Intuitively, this is expected, because the former pays the price of higher complexity due to the use of integer programming.
In contrast, the latter is a polynomial time algorithm which is useful for large-scale scenarios.
Therefore, the MAUU algorithm illustrates excellent tradeoff between complexity and accuracy.

\section{Conclusions} \label{Conclusions}
This paper has investigated the caching problem with presence of user mobility, for which the inter-contact model is used
to describe the mobility pattern of mobile users. An optimization problem, COCP, has been modelled, analyzed and formulated.
The hardness of the problem has been thoroughly proved via a reduction from the 3-SAT problem. For problem-solving, two computational approaches, namely, the ACOCP approach and the MAUU algorithm, have been developed.
Performance evaluation shows that the two approaches result in significant improvement in comparison to conventional caching algorithms.
Moreover, solving ACOCP leads to an effective approximation scheme, and the MAUU algorithm achieves excellent balance between complexity and accuracy.

An extension of the work is the consideration of a more complicated hierarchical caching architecture with presence of mobility, i.e., caching at both users and base stations. This can be formulated as to minimize the expected delay for recovering one file, with constraints on the total number of encoded segments and cache capacity.

\begin{appendices}
\section{}\label{Appex_Lower}
To facilitate presentation, define
\begin{equation}
\begin{aligned}
\Delta^\text{n1}_{fi}\triangleq\text{E}\{\max[S^f_\text{rec}-(\underset{j\in \mathcal{U},j\neq i}\sum \min(BM_{ij},x_{fj})+x_{fi}),0]\},
\label{X1}
\end{aligned}
\end{equation}
and
\begin{equation}
\Delta^\text{n2}_{fi}\triangleq\left\{
\begin{aligned}
{}&S^f_\text{rec}-[\text{E}(\underset{j\in \mathcal{U},j\neq i}\sum \min(BM_{ij},x_{fj}))+x_{fi} ],\\
&~\text{if}~S^f_\text{rec}>\text{E}(\underset{j\in \mathcal{U},j\neq i}\sum \min(BM_{ij},x_{fj}))+x_{fi}, \\
{}&0,~\text{else}.
\label{X3}
\end{aligned}
\right.
\end{equation}
Given $\bm{x}$, we will prove the relationship between $\Delta^\text{n1}_{fi}$ and $\Delta^\text{n2}_{fi}$.

(i) When $S^f_\text{rec}>\text{E}(\underset{j\in \mathcal{U},j\neq i}\sum \min(BM_{ij},x_{fj}))+x_{fi} $, it follows that
\begin{equation}
\begin{aligned}
&S^f_\text{rec}-[\text{E}(\underset{j\in \mathcal{U},j\neq i}\sum \min(BM_{ij},x_{fj}))+x_{fi}] \\
= & \text{E}[S^f_\text{rec}-(\underset{j\in \mathcal{U},j\neq i}\sum \min(BM_{ij},x_{fj})+x_{fi}) ].
\label{X4}
\end{aligned}
\end{equation}
Due to the fact that
\begin{equation}
\begin{aligned}
&S^f_\text{rec}-(\underset{j\in \mathcal{U},j\neq i}\sum \min(BM_{ij},x_{fj})+x_{fi}) \\
\le &\max[ S^f_\text{rec}-(\underset{j\in \mathcal{U},j\neq i}\sum \min(BM_{ij},x_{fj})+x_{fi}),0 ],
\label{X5}
\end{aligned}
\end{equation}
it follows that
\begin{equation}
\begin{aligned}
&\text{E}\{S^f_\text{rec}-(\underset{j\in \mathcal{U},j\neq i}\sum \min(BM_{ij},x_{fj})+x_{fi})\} \\
\le & \text{E}\{\max[ S^f_\text{rec}-(\underset{j\in \mathcal{U},j\neq i}\sum \min(BM_{ij},x_{fj})+x_{fi}),0 ]\}.
\label{X6}
\end{aligned}
\end{equation}
Combining (\ref{X4}), (\ref{X5}), with (\ref{X6}), we obtain
\begin{equation}
\begin{aligned}
&S^f_\text{rec}-[\text{E}(\underset{j\in \mathcal{U},j\neq i}\sum \min(BM_{ij},x_{fj}))+x_{fi} ]\\ \le & \text{E}\{\max[ S^f_\text{rec}-(\underset{j\in \mathcal{U},j\neq i}\sum \min(BM_{ij},x_{fj})+x_{fi}),0 ]\}.
\label{X7}
\end{aligned}
\end{equation}
Thus, $\Delta^\text{n1}_{fi}\ge \Delta^\text{n2}_{fi}$.

(ii) When $S^f_\text{rec}\le \text{E}(\underset{j\in \mathcal{U},j\neq i}\sum \min(BM_{ij},x_{fj}))+x_{fi} $, $\Delta^\text{n2}_{fi}=0$.
Assume that $m$ experiments are conducted.
Denote by $S^r_{fi}$ the number of segments of file $f$ collected by user $i$ in the $r$th experiment, $r=1,2,\dots,m$, and $S^f_\text{rec}\le \frac{1}{m}\sum_{r=1}^{m}S^r_{fi}$. There are two cases. The first case is that user $i$ can successfully recover the file $f$ in each experiment, i.e., $S^r_{fi}\ge S^f_\text{rec}, r=1,2,\dots,m$. In this case, $\Delta^\text{n1}_{fi}= \Delta^\text{n2}_{fi}$. The second case is that user $i$ unsuccessfully recovers the file $f$ at least one experiment. For the second case, $\Delta^\text{n1}_{fi}>0$. Thus, $\Delta^\text{n1}_{fi}> \Delta^\text{n2}_{fi}$.

Combining (i) with (ii), it follows that
\begin{equation}
\begin{aligned}
\Delta^\text{n1}_{fi}\ge \Delta^\text{n2}_{fi}.
\label{X2}
\end{aligned}
\end{equation}
Therefore, $\Delta \ge  \Delta^{lb}$.

 \section{} \label{Appex_iteration}
$ \Delta$ can be simplified as
 \begin{equation}
\begin{aligned}
  \Delta
=&\frac{1}{U}\underset{i\in \mathcal{U}}\sum \underset{f\in \mathcal{F}}\sum P_{fi}[ \underset{j\in \mathcal{U},j\neq i}\sum \text{E}(S_{fi}-x_{fi})\delta_\text{D}\\
&~~+ \sum_{k=x_{fi}}^{S^f_\text{rec}-1}(S^f_\text{rec}-k)\text{Pr}(S_{fi}=k) \delta_\text{N}   ],
\label{SimObjeFun}
\end{aligned}
\end{equation}
where
\[
\text{Pr}(S_{fi}=k)=\text{Pr}(\underset{j\in \mathcal{U},j\neq i}\sum \min(BM_{ij},x_{fj})+x_{fi}=k).
\]
Computing $\text{Pr}(S_{fi}=k)$ directly by using multiple summations, the computational complexity exponentially increases with $U $. However, we can use a recursive function with polynomial-time complexity. Define
\begin{equation}
\begin{aligned}
\text{Pr}(U,k)&\triangleq \text{Pr}(S_{fi}=k) \\
&=\text{Pr}(\underset{j\in \mathcal{U},j\neq i}\sum \min(BM_{ij},x_{fj})+x_{fi}=k).
\end{aligned}
\end{equation}
 After some mathematical manipulations, the recursive function can reformulated as
 \begin{equation}
\begin{aligned}
\text{Pr}(U,k)=&\sum_{t=0}^{k}[\text{Pr}(\min(BM_{i,U},x_{f,U})=t) \\
&~~~*\text{Pr}(U-1,k-t)].
\end{aligned}
\end{equation}
The above function manifests that if $t$ segments are collected from user $U$, then user $i$ will obtain $k-t$ segments from the other $U-1$ users including itself. In general, $\text{Pr}(U-\tau,.)$ depends on $\text{Pr}(U-\tau-1,.)$, $\tau=0,1,\dots,U-2$, leading to a recursive process.
The overall complexity of computing $ \Delta$ is $O(FU^2S{'}^2_\text{rec})$, where $S{'}_\text{rec}=\underset{j \in \mathcal{F}}{\text{max}}~S^j_\text{rec}$.

 \end{appendices}

\bibliographystyle{IEEEtran}


\end{document}